\documentclass[aps,pra,twocolumn,twoside,superscriptaddress,notitlepage]{revtex4-2}
\usepackage{tikz}
\usetikzlibrary{calc}
\usetikzlibrary{shapes.geometric, arrows}
\usepackage{eufrak}
\usepackage{enumitem}
\usepackage{mathtools}
\usepackage{graphicx}
\usetikzlibrary{positioning}
\usepackage{bm}
\newcommand{\bematrix}{\left(\begin{matrix}}
\newcommand{\ematrix}{\end{matrix}\right)}

\usepackage{ulem} % only here for strikethrough-font
\usepackage[caption=false]{subfig}
\usepackage{dcolumn}
\usepackage{amsmath,amssymb}
\usepackage{bm}
\usepackage{bbm}
\usepackage{overpic}
\usepackage{latexsym}
\usepackage{color}
\usepackage[english]{babel}
\usepackage{latexsym}
\usepackage{psfrag,graphicx}
\usepackage{epsf}
\usepackage{amsmath}
\usepackage{amssymb}
\usepackage{amsfonts}
\usepackage{natbib}
\usepackage{multirow} 
\usepackage{appendix}
\usepackage{verbatim}
\usepackage{enumitem}
\usepackage{dsfont}
\usepackage{float}
\usepackage{tikz}

\newcommand{\T}{\text{T}}
\definecolor{mygrey}{gray}{0.35}
\definecolor{myblue}{rgb}{0.2,0.2,0.8}
\definecolor{myzard}{cmyk}{0,0,0.05,0}
\definecolor{mywhite}{rgb}{1,1,1}
\definecolor{myred}{rgb}{0.9,0.1,0.}
\usepackage[colorlinks=true,citecolor=myblue,linkcolor=myblue,urlcolor=myblue]{hyperref}

\usepackage[makeroom]{cancel}

\newenvironment{proof-of}[1]{\medskip\noindent\textbf{Proof of {#1}.}}{\hfill$\blacksquare$\medskip}
\newcommand{\ket}[1]{\left\vert#1\right\rangle}

\newcommand{\braket}[2]{\ensuremath{\langle #1 | #2 \rangle}}

\usepackage{colortbl}
\usepackage{xcolor}

\usepackage{tcolorbox}

\definecolor{lightgray}{gray}{0.9}

\begin{document}

\title{Information geometric quantification of effective privacy in quantum metrology}

\author{Luca Bianchi}
\affiliation{Department of Physics and Astronomy, University of Florence, 50019, Firenze, Italy}
\email{luca.bianchi@unifi.it}

\author{Shimpei Yamaguchi}
\affiliation{Department of Electronics and Electrical Engineering, Keio University, 3-14-1 Hiyoshi, Kohoku-hu, Yokohama 223-8522, Japan}

\author{Wojciech Roga}
\affiliation{Department of Electronics and Electrical Engineering, Keio University, 3-14-1 Hiyoshi, Kohoku-hu, Yokohama 223-8522, Japan}

\author{Davide Bacco}
\affiliation{Department of Physics and Astronomy, University of Florence, 50019, Firenze, Italy}

\author{Masahiro Takeoka}
\affiliation{Department of Electronics and Electrical Engineering, Keio University, 3-14-1 Hiyoshi, Kohoku-hu, Yokohama 223-8522, Japan}

\begin{abstract}
    Privacy of a quantum metrological protocol concerns the extent to which single parameters can be kept inaccessible to an observer or to other users of the network. 
    In this work, an information geometric framework is developed to quantify privacy and accessibility of functions of parameters effectively, that is, up to a finite accuracy in state discrimination. 
    Both quantities are defined by measuring volumes in the parameter space induced by the underlying quantum states. 
    This construction subsumes previous definitions of privacy based on the degeneracy of quantum Fisher information, naturally encompassing imperfect implementations. 
    Using extended-GHZ states as a representative example of a quantum network scenario, privacy and accessibility are characterized by quantum correlations and accuracy, providing scaling laws depending on imperfect measurements and entanglement.  
\end{abstract}

\date{\today}

\maketitle

\section{Introduction}

In the last decades, quantum information processing has reached an unprecedented development, controlling and leveraging unique quantum-mechanical features ---such as superposition and entanglement \cite{horodecki2009quantum}--- to enhance computation \cite{shor1999polynomial}, communication \cite{ekert1991quantum}, and estimation \cite{giovannetti2011advances} of information in ways that are impossible within classical physics.

Among these developments, quantum networks \cite{azuma2023quantum} have attracted significant attention as architectures that distribute quantum resources across spatially separated nodes. 
In such networks, entanglement and shared quantum states enable coordinated tasks including distributed computation \cite{barral2025review}, clock synchronization \cite{komar2014quantum,azahari2024review} and enhanced metrology \cite{degen2017quantum}. 
A particularly important feature of quantum networks is parameter sharing, where multiple nodes jointly estimate or encode physical parameters using correlated quantum probes \cite{zhang2021distributed}. 
By exploiting multipartite entanglement and collective measurements, quantum-enhanced networks can surpass the precision limits achievable by independent sensors \cite{toth2014quantum}.

As quantum networks become of increasing importance, privacy in distributed quantum sensing has emerged as an active research direction, driven by the need to protect individual data while still enabling collective parameter estimation. 
Early foundational work established protocols for private network parameter estimation using GHZ states with quantum state verification \cite{shettell2022private}, and demonstrated that secure sensing is achievable even without entanglement \cite{moore2023secure}. 
The cryptographic approach to quantum 
metrology has been explored both theoretically \cite{huang2019cryptographic,shettell2022cryptographic} and experimentally \cite{yin2020experimental, ho2026quantum}, providing rigorous adversarial security guarantees. 
Complementary to this, anonymity and integrity in quantum sensor networks have been addressed through anonymous parameter estimation protocols \cite{de2025anonymous}, while secure sensing against man-in-the-middle attacks has been proposed using GHZ and decoy-state strategies \cite{moore2025secure}.

More recently, a framework for privacy \cite{hassani2025privacy} was defined operationally in terms of quantum Fisher information (QFI), a central object of interest in quantum metrology.
Subsequent work has analyzed the robustness of private states against noise \cite{bugalho2025private} and extended the framework to continuous-variable and Gaussian settings \cite{junior2025privacy,alushi2026privacy}. 
The operational privacy framework based on the classical Fisher information matrix was further developed in \cite{namkung2026universal}, while the composability of privacy across networked sensing scenarios has been investigated in \cite{solomons2025composable}. 

\begin{figure}
    \centering
    \includegraphics[width=\linewidth]{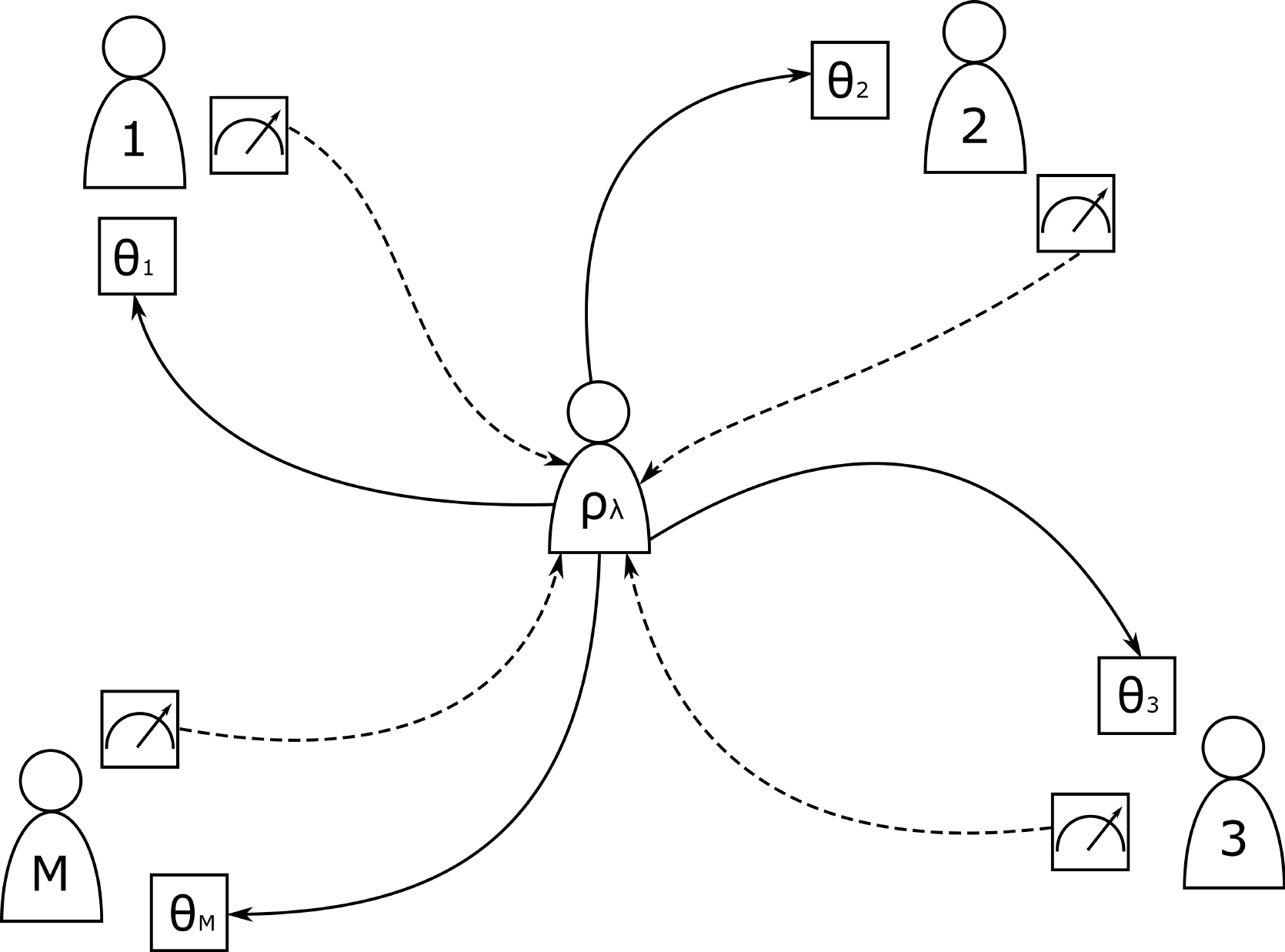}
    \caption{
    A typical quantum network sensing protocol:
    The central node prepares a multipartite entangled state $\rho_{\lambda}$, sharing with $M$ parties across quantum channels (solid lines) in the network.
    Each node encodes a parameter $\theta_i$, with $i=\{1,\dots,M\}$ and performs a measurement. 
    Measurement statistics are then classically communicated (dashed lines) back to the central node, who builds estimators on top of the received probabilities.
    Such estimators can have high or low variances depending on the quantum Fisher information of the multipartite encoded quantum state $\rho_{\lambda,\theta}$ measured.
    }
    \label{fig:quantum_network}
\end{figure}

In a typical quantum network metrology, a quantum state is distributed across spatially separated users of the network, who imprint a parameter $\theta_i$ through a local unitary $U_i$ and subsequently perform a measurement. (See fig. \ref{fig:quantum_network}.)
By means of the measurement statistics $p_i$ that are classically communicated to the central node, estimators $\hat\theta_i$ can be built to infer parameters and their functions.
The variance of such estimators are governed by the quantum Cramer-Rao bound \cite{braunstein1994statistical}, viz. $\Delta\theta_{ij} \succ q^{+}_{ij}/\mathcal{N}$, with $q^{+}_{ij}$ the Moore-Penrose pseudo-inverse of the QFI $q_{ij}$ \cite{helstrom1969quantum}, ``$\succ$" the Loewner order and $\mathcal{N}$ the number of repetition of the measurements. 
The QFI therefore quantifies the sensitivity of the network to variation of parameters.
However, it also provides a natural metric in the context of information geometry \cite{amari2016information, lambert2023classical}, where distinguishability between quantum states is computed by means of the Bures distance, which is later defined. 
Here, quantum states are points on a projective manifold equipped with a metric tensor \cite{bengtsson2017geometry} and concepts such as superposition and entanglement are naturally accommodated to have a geometrical meaning \cite{brody2001geometric}.

The aim of this work is to establish the theory of privacy in the context of quantum information geometry, in order to give a quantification of privacy provided by quantum states. 
Privacy and accessibility are quantified in terms of the number of encoded sets of parameters associated to a quantum state distinguishable up to a given accuracy.
The latter naturally provides for a pragmatic characterization of privacy in terms of computations of measures. 
To this end, quantum states are regarded as linear surjective maps between the space of parameters and the Hilbert space, treated both as differentiable manifolds. 
Computing privacy then amounts to compute volumes induced by the push-forward Bures distance on the space of parameters, restricting to foliations of the latter whenever the Fisher information is degenerate, thus subsuming previous approaches \cite{hassani2025privacy}.
Furthermore, in the geometric approach, special attention is payed to quantification of the effective privacy achievable with limited resources like the number of repetitions, the accuracy of measuring devices, and the quantum state preparation imperfections, paving the way for simulating practical implementation of private metrological protocols.

\section{Methods}
\label{sec:methods}
\subsection{Preliminaries}
\label{subsec:preliminaries}
\begin{figure}
    \centering
    \includegraphics[width=\linewidth]{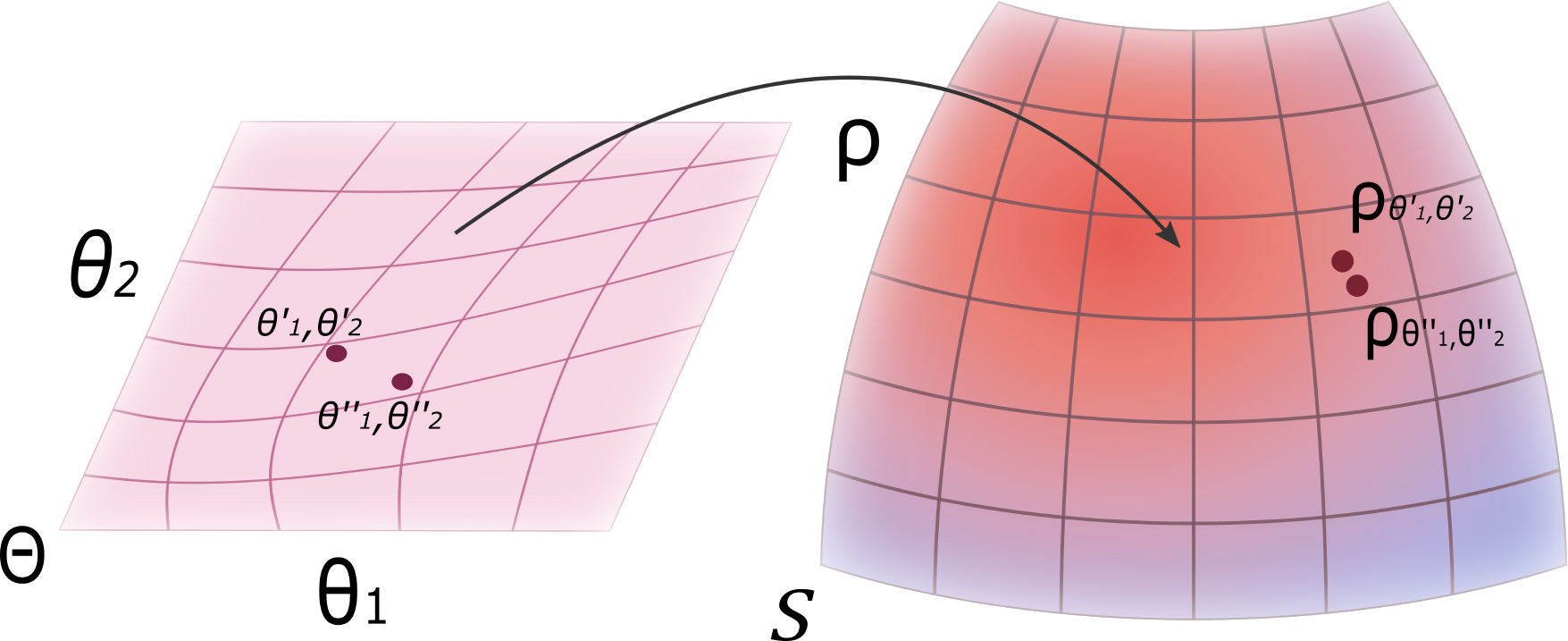}
    \caption{Schematic picture for the encoding of parameters in quantum states:
    on the left, the space of parameters $\Theta$ while on the right the space of states $\mathcal S$, the two spaces connected by the mapping $\rho$.
    Colors stand for measures in the spaces, while grids denote distinguishability regions.
    In this example, the space of parameters has a uniform measure, while the space of parametrized quantum states has an induced non-uniform density, portrayed as a gradient of colors.
    In the quantum state manifold, a uniform grid indicates region of distinguishability standing for volumes of $\epsilon$-balls, which are mapped back in twisted, generally-irregular grids on the space of parameters.
    }
    \label{fig:manifold}
\end{figure}

Quantum states, being constrained by normalization and being equivalent up to a global phase, make the Hilbert space a complex projective space \cite{brody2001geometric}, which can be regarded as a K\"ahler manifold \cite{silva2001lectures}, denoted here as $\mathcal S$. 
(See appendix \ref{appendix:dif_geo} for an introduction to core concepts of differential geometry.)
In this picture, if one introduces continuous parameters $\{\theta_i\}_{i=1}^M$ belonging to a space $\Theta$, each state $\rho$ can be regarded as a map $\rho:\Theta \to \mathcal S$ such that $\rho(\theta_1,\dots,\theta_M) :=\rho_{\theta} \in \mathcal{S}$.
The relation between the space of parameters and the manifold of quantum states is qualitatively shown in Fig. \ref{fig:manifold}.
Furthermore, one can locally define tangent vectors $\dot\rho(t) = d\rho/dt$ for each value of a real variable $t$ parametrizing a trajectory in the Hilbert space.
Therefore, each point on the manifold is equipped with a vector space, called tangent space, where an appropriate basis can be introduced.
In particular, it is customary to chose a coordinate basis given by $\{\partial_{\theta_i}\}_{i=1}^M$. 
The K\"ahler structure of the manifold naturally introduces the concept of a metric tensor, denoted by $g$, which turns out to be the Bures metric \cite{bures1969extension}. 
Feeding the quantum state $\rho$ to the Bures metric amounts to apply the \emph{pull-back} $\rho^*$ to the tensor $g$, i.e. $\rho^*[g]$.
This, by definition, equivalently amounts to evaluate the Bures metric on the tangent vectors obtained by the \emph{push-forward} $\rho_*$, i.e. in coordinate basis
\begin{equation}
    \rho^*[g]_{ij} 
    :=
    g(\rho_*[\partial_{\theta_i}], \rho_*[\partial_{\theta_j}])
    =
    g(\partial_{\theta_i}\rho_{\theta},\partial_{\theta_j}\rho_\theta) 
\end{equation}
which is, in turn
\begin{equation}
    \label{def:bures_metric}
        g(\partial_{\theta_i},\partial_{\theta_j})
        = 
        \frac{1}{2}
        \mathrm{Re}
            \left[
                \sum_{k,l=1}^N
                    \frac{
                        \braket{
                            k|\partial_{\theta_i}\rho_{\theta}}{l
                            }
                        \braket{
                            k|\partial_{\theta_j}\rho_{\theta}}{l}
                        }
                    {p_k + p_l}
            \right].
\end{equation}

Here, $k,l$ are indices for elements of orthonormal eigenvectors of $\rho_{\theta}$, each one with probability $p_k$. 
Notice that the above metric is just one fourth of the QFI $q_{ij}$, thereby governing the distinguishability in terms of distance between quantum states locally.
As was proven by Uhlmann, the fidelity between two quantum states $\rho_{\theta},\rho_{\theta'}$ is related to the geodesic distance defined by the Bures metric, thereby giving the Bures distance \cite{wootters1981statistical}:

\begin{equation}
    \label{def:bures_distance}
        D_B(\rho_{\theta},\rho_{\theta'}) 
        =
        \arccos{
            \left[
                \sqrt{
                        F(
                            \rho_{\theta},\rho_{\theta'}
                        )
                    }
            \right]
        },
\end{equation}
with $F(\rho_{\theta},\rho_{\theta'})$ the fidelity between the quantum states $\rho_{\theta},\rho_{\theta'}$. 
For pure states, the latter reduces to 
$
    F(\rho_{\theta},\rho_{\theta'}) 
    = 
    |
        \braket{\Psi_{\theta}}{\Psi_{\theta'}}
    |^2.
$
The Bures distance will be adopted in order to characterize neighborhoods in the space of quantum states $\mathcal{S}$.

\subsection{Privacy and accessibility}
\label{subsec:definitions}

The general concept of privacy of parameters encoded in quantum states is related to the distinguishability between these states. 
One says that a set of parameters encoded in quantum states remains private if we cannot tell which selection from this set was encoded. 
This requires the states corresponding to parameters chosen from the private set to be practically indistinguishable. 
This condition bounds their trace distance and, consequently, limits the optimal success probability for discriminating between them according to Helstrom's theory of quantum state discrimination. 
It is therefore natural to regard all states within a Bures ball of radius $\epsilon\in[0,\pi/2]$ 
\begin{equation}
    \label{def:epsilonball}
        B_\epsilon(\rho)
        =
        \{
            \rho':D(\rho,\rho')\leq \epsilon
        \}
\end{equation}
as operationally equivalent at the $\epsilon$-level.
On the other hand, two choices of parameters are accessible at scale $\epsilon$ if the Bures distance between the corresponding quantum states is larger than $\epsilon$.
In what follows, $\epsilon$ can be regarded as a measure of accuracy, in the state discrimination sense, therefore connecting to the Chernoff theory of error exponents \cite{audenaert2007discriminating}.
(See Appendix \ref{appendix:distinguishability} for details.)

Consider the measure of the $\epsilon$-ball in the set of quantum states
\begin{equation}
    \label{def:epsilonvolume}
        {\mathcal M}_\epsilon(\rho)
        =
        \int d\mu(\rho')
            H[
            \epsilon-D(\rho,\rho')
                ]
\end{equation}
with $\mu(\rho')$ a given measure on the space of states, $H[x-y]$ the Heaviside function, given by
\begin{equation}
    H[x-y] 
    = 
    \begin{cases}
        1 \;\; x\geq y \\ 
        0 \;\; x < y
    \end{cases}\;,
\end{equation}
and $\epsilon$ a parameter determining the size of the region. 
Let the measure of the total space be 
\begin{equation}
{\cal M}=\int d\mu(\rho).
\end{equation}
The elementary measure $d\mu(\rho)/{\mathcal M}$ is the probability that a randomly sampled state is in this element of volume.

Because the joint probability of two independent events is the product of their respective probabilities, 
the probability that two 
independent and identically distributed (\emph{i.i.d.}) randomly chosen states belong to the same element of volume 
$
d\mu(\rho)
$
is
$(d\mu(\rho)/{\mathcal M})^2$. 
Therefore, the probability that two randomly sampled states are in the same volume element is
\begin{equation}
    p
    =
    \frac{1}
    {{\cal M}^2}\int d\mu(\rho)
     d\mu(\rho),
    \label{eq:renyi}
\end{equation}
while the probability that two randomly sampled states are $\epsilon$-indistinguishable is
\begin{equation}
\label{def:epsilon_probability}
    p_{\epsilon}
    =
    \frac{1}{{\cal M}^2}\int d\mu(\rho){\mathcal M}_\epsilon(\rho),
\end{equation}
which just requires that both states are sufficiently $\epsilon$-close.

From the probability $p_{\epsilon}$, one can obtain the ratio of the total measure of states to the average measure of an $\epsilon$-ball
\begin{equation}
    \label{def:accessibility}
    A_\epsilon
    =
    \frac{\cal M}
    {
        \frac{1}{\cal M}
        \int d\mu(\rho)
            {\mathcal M}_\epsilon(\rho)
    }.
\end{equation}
The above quantity can be interpreted as the effective number of states that can be distinguished up to an $\epsilon$ accuracy, thereby formally defining an $\epsilon$-\emph{accessibility} function. 
If the number of distinguishable states is large, or if the probability $p_{\epsilon}$ is high, the  accessibility $A_\epsilon$ is also large.
The logarithm of $A_\epsilon$ is known as the generalized Rényi 2-entropy at scale $\epsilon$  \cite{luczak1997suboptimal, seleznjev2010random,leonenko2010statistical}. 
Furthermore, the above concept nicely connects with the $\epsilon$-variability \cite{seleznjev2010random}. 
The quantity  $A_{\epsilon}$ evaluated for $\epsilon \rightarrow 0$ is called the inverse participation ratio studied in the context of random matrix theory \cite{zyczkowski2003renyi}.
Moreover, the generalized Rényi-entropy for the discrete uniform distribution corresponds to the Kolmogorov entropy \cite{seleznjev2010random,leonenko2010statistical}, which describes the minimum cover of the space by $\epsilon$-balls.
The standard Renyi 2-entropy \cite{renyi1961measures} is recovered in the limit of small $\epsilon$. 
The latter is sometimes called the collision entropy \cite{bosyk2012collision, ribeiro2021entropy}, and also applied as a sparsity measure \cite{wei2024investigations}, that is, how low is on average the density of points on the space.
All the aforementioned cases use formulae analogous to \eqref{def:accessibility} to characterize different kinds of distinguishability of variables in their space, which is the effective number of different sets of elements that can be distinguished (assuming $\epsilon$ accuracy). 

The formal definition of $\epsilon$-accessibility \eqref{def:accessibility} allows to formalize the concept of $\epsilon$-level \emph{privacy}.
The latter is defined as the average volume of the states that cannot be distinguished at $\epsilon$-scale, which coincides with the measure of the $\epsilon$-ball
\begin{equation}
    \label{def:privacy}
        P_\epsilon
        =
        \frac{1}{\mathcal{M}^2}
        \int d\mu(\rho)
            {\mathcal M}_\epsilon(\rho).
\end{equation}
This quantity characterizes the average amount of states concentrated in any ${\cal M}_{\epsilon}(\rho)$. 
Intuitively, if such quantity is large, many states are more densely concentrated, making them hard to distinguish.
By contrast, a small $P_\epsilon$ indicates no special concentration of states in the $\epsilon$-ball and high probability of distinguishability of two randomly selected states.
It is straightforward to check that accessibility and privacy satisfy a duality relation 
$$
A_\epsilon P_\epsilon= 1,
$$
pointing out a trade-off between distinguishability of states and access to parameters.
The interpretation of $\log A_{\epsilon}$ as an entropy helps the intuition that accessibility describes the amount of information about the precise set of parameters involved, that is, about the number of bits accessible through measurement of quantum states. 
Given that the full knowledge of the collection of parameters requires a given number of bits, $\epsilon$-accessibility indicates also how many bits are missing or hidden in the quantum noise and assumed imperfections. 
However, the accessibility measure does not specify which piece of information is missing. 
The latter is the subject of the following section.

%The measures of $\epsilon$-privacy and $\epsilon$-accessibility are invariant under unitary transformation in the space of states, which is the consequence of invariance of the Bures angle distance. 
%Also, linear and reversible transformations in the space of parameters leave the measures invariant. \W{[Proof needed!]} These are natural expectation from reasonable measures of both privacy and accessibility. They should not change under trivial transformations like renaming parameters or changing the basis of description of the quantum states.

\subsection{Geometric privacy and accessibility}
\label{subsec:geometric_pa}

Given a measure $\mu_{\theta}$ on the parameter space $\Theta$ and a measurable region $R$ in the state space $\mathcal S$, the state map $\rho$
induces a measure on the state manifold through the push-forward construction
\begin{equation}
    \label{def:pushforward}
    (\rho_{\ast}\mu_{\theta})(R)
    =
    \mu_{\theta}\!\left(
        \rho^{-1}(R)
    \right) := \mu(\rho_\theta).
\end{equation}
In other words, the measure of a set of states is defined as the measure of all parameter values producing those states.

When the pull-back of the Bures metric to the parameter space is non-degenerate, the corresponding volume form is uniquely determined by the metric tensor $g$. 
In local coordinates,
\begin{equation}
    \label{def:volumeform}
    d\mu(\rho_\theta)
    =
    \sqrt{
\det[g_\theta]
    }
    \,
    d\theta,
\end{equation}
where $d\theta=d\theta_1\cdots d\theta_M$.

The geometry of quantum states naturally provides practical tools to compute $\epsilon$-accessibility and $\epsilon$-privacy through
\begin{equation}
    \label{def:integral_formula}
    {\mathcal M}_\epsilon(\rho_{\theta})
    =
    \int
    d\theta'
    \sqrt{ \det[g_{\theta'}]
    }
    H
        \left[
            \epsilon
            -
            D_B(
                \rho_{\theta},
                \rho_{\theta'}
            )
        \right].
\end{equation}
The total accessible volume is correspondingly
\begin{equation}
    \mathcal M
    =
    \int
    d\theta
    \sqrt{
        \det[g_\theta]
    }
.
\end{equation}

From the definitions \eqref{def:accessibility} and \eqref{def:privacy} one can naturally establish a quantification of the previously defined concept of privacy \cite{hassani2025privacy}, which is referred to here as \emph{directional privacy} or \emph{local privacy}. 
In this framework, functions of the parameters can be hidden along unobservable directions, identified with tangent vectors belonging to the kernel of the QFI, $\mathrm{Ker}[g]$.
This construction is inherently local, since it relies on the tangent space attached to each point of the manifold.

Suppose that the QFI admits $p$ independent null directions, while the remaining $N=M-p$ directions belong to its complement. 
If the eigenvectors of the QFI define a coordinate basis, one may introduce new coordinates
\begin{equation}
    \{
        \alpha_a
    \}_{a=1}^{N},
    \qquad
    \{
        \beta_b
    \}_{b=1}^{p},
\end{equation}
where the observable coordinates $\{\alpha_a\}_{a=1}^N$ span the complement of $\mathrm{Ker}[g]$, while the coordinates $\{\beta_b\}_{b=1}^p$ parameterize the hidden directions. 
A practical scenario happens when, for example, a quantum state depends only on the coordinates $\alpha_a$, whereas motion along the $\beta_b$ directions leaves the state unchanged.
The Jacobian
\begin{equation}
    J_{a}^{\;\;i}
    =
    \frac{
        \partial \theta_i
    }{
        \partial \alpha_a
    }
\end{equation}
induces on the observable subspace the restricted metric
\begin{equation}
    \label{eq:restriction}
    h_{ab}
    =
    \sum_{i,k=1}^{M}
    J_{a}^{\;\;i}
    g_{ik}
    J_{b}^{\;\;k}.
\end{equation}
By construction, $h$ is non-degenerate and contains all information accessible through the quantum state manifold.

The condition $\det[g]=0$ implies that the volume form \eqref{def:volumeform} is no longer defined on the full parameter space. 
In this case, one must separate observable and hidden directions. 
The resulting measure can be written, up to an overall normalization factor, as
\begin{equation}
    \label{eq:coarea_measure}
    d\mu(\rho)
    =
    \sqrt{\det[h_\alpha]
    }
    d\alpha
    d\beta,
\end{equation}
where
\begin{equation}
    d\alpha
    =
    d\alpha_1
    \dots
    d\alpha_N,
    \qquad
    d\beta
    =
    d\beta_1
    \dots
    d\beta_p.
\end{equation}
The factor $\sqrt{\det[h_{\alpha}]}$ quantifies distinguishability in the space of quantum states, while the coordinates $\beta$ contribute through the measure inherited from the original parameter space. 
Geometrically, all parameter values sharing the same observable coordinates $\alpha$ correspond to the same quantum state and are therefore counted through the integration over $\beta$.

Privacy measures are then computed through
\begin{equation}
    \mathcal M_{\epsilon}
    (\rho_{\alpha})
    =
    \int
    d\beta'
    \int
    d\alpha'
    \sqrt{
        \det[
            h_{\alpha'}
        ]
    }
    H
    \left[
        \epsilon
        -
        D_B(
            \rho_{\alpha},
            \rho_{\alpha'}
        )
    \right],
\end{equation}
which is a particular instance of the coarea formula \cite{federer1959curvature}.
The total measure becomes
\begin{equation}
    \mathcal M
    =
    \int
    d\beta
    \int
    d\alpha
    \sqrt{\det[h_{\alpha}]}
    \mathcal M_{\epsilon}
    (\rho_{\alpha}).
\end{equation}

As a simple example, consider the Bell state
\begin{equation}
    \ket{\psi}
    =
    \frac{1}{\sqrt 2}
    \left(
        \ket{00}
        +
        \ket{11}
    \right),
\end{equation}
with parameters $(\theta_1,\theta_2)$ encoded through local phase rotations,
\begin{equation}
    \ket{\psi_{\theta}}
    =
    \frac{1}{\sqrt 2}
    \left(
        \ket{00}
        +
        e^{i(\theta_1+\theta_2)}
        \ket{11}
    \right).
\end{equation}
The parameter space is a two-dimensional torus, whereas the quantum state depends only on the combination
\begin{equation}
    \alpha
    =
    \theta_1+\theta_2.
\end{equation}
The orthogonal combination
\begin{equation}
    \beta
    =
    \theta_1-\theta_2
\end{equation}
does not affect the state and therefore corresponds to a private direction. This is reflected by the pull-back of the Bures metric onto the parameter manifold, which is singular. Restricting the metric to the observable coordinate $\alpha$ yields a one-dimensional non-degenerate metric $h$. 
Consequently, the volume element takes the form
\begin{equation}
    d\mu(\rho)
    =
    \frac{1}{2}
    \sqrt{h}
    d\alpha
    d\beta,
\end{equation}
and the integration over $\beta\in[-2\pi,2\pi)$ counts all parameter values corresponding to the same physical state.

\section{Applications}
\label{sec:applications}
In order to exemplify the method discussed in the previous section, consider the case of a quantum network of sensors (see Fig.~ \ref{fig:quantum_network}).
The initial state of the central node of the network can be taken from a family of extended $(\lambda,\gamma)$-GHZ states:
\begin{equation}\label{def:family_maintext}
            \ket{\Psi_{\gamma, \lambda,\theta}} 
            = 
            A_{\lambda, \gamma}
            \ket{0}^{\otimes M}
            + 
            B_{\lambda,\gamma,\theta}
            \ket{1}^{\otimes M}
            +
            \sum_{x\in \mathcal{P}_M}
                C_{x, \lambda,\gamma,\theta}
                \ket{x}
    \end{equation}
     where $\mathcal{P}_M = \{0,1\}^{M}\setminus\{0^M,1^M\}$,
  \small{
    \begin{equation}
        \begin{cases}
            A_{\lambda, \gamma} = \sqrt{2^M\lambda^2\,\gamma^2+(1-2^M\lambda^2)\,\gamma^{2M}} \\
            B_{\lambda,\gamma,\theta} = \sqrt{2^M\lambda^2(1-\gamma^2)+(1-2^M\lambda^2)(1-\gamma^2)^M}
                e^{i\sum_{j=1}^M\theta_j} \\
            C_{x,\lambda, \gamma,\theta} = \sqrt{1-2^M\lambda^2}\,\gamma^{M-|x|}(1-\gamma^2)^{|x|/2}
                e^{i\sum_{j=1}^M x_j\theta_j}\\
        \end{cases}
    \end{equation}
    }
    and $|x| = \sum_j x_j$ is the Hamming weight.
(See appendix \ref{appendixA:application_details} for definitions, details and derivations of the results given below.)

The set of parameters 
$\{\theta_i\}_{i=1}^M$ is encoded by the users of the network, while the parameter $\lambda$ effectively denotes the amount of entanglement.
At $\lambda=0$ one gets a separable state, while at $\lambda=2^{-M/2}$ one obtains the $\gamma$-GHZ-like state, the usual GHZ state recovered for $\gamma=1/\sqrt{2}$.
Even though the parameters can be treated on equal footing from the point of view of quantum sensing theory, the only interest here regards the privacy and the accessibility of the $\theta_i$ parameters.
For this state, the QFI defines a flat geometry in the space of parameters $\theta_i$
\begin{equation}
        q_{ij} 
        = 
        4\gamma^2(1-\gamma^2)
            \left[
                (1-2^M\lambda^2)
                \mathrm{id}_M 
                + 
                2^M\lambda^2
                \mathrm{J}_M
            \right],
    \end{equation}
with $\mathrm{J}_M = \mathbf{1}\mathbf{1}^{\T}$ and $\mathbf{1}^{\T}$ the $M$-dimensional vector $\mathbf{1}^{\T} = (1,\dots,1)$.
Its spectrum has an eigenvalue $\mu^+ = 4\gamma^2(1-\gamma^2)[1+(M-1)2^M\lambda^2]$, with multiplicity one corresponding to the eigenvector $w^+ = (1/\sqrt{M})\mathbf{1}$ and an eigenvalue $\mu^- = 4\gamma^2(1-\gamma^2)[1-2^M\lambda^2]$ with multiplicity $M-1$, related to the eigenspace spanned by $w^k = e_1 - e_{k+1}$. 
($e_k$ being the $k$-th vector of the canonical basis of the space of parameters.)

\begin{figure*}
    \centering
    \includegraphics[width=\linewidth]{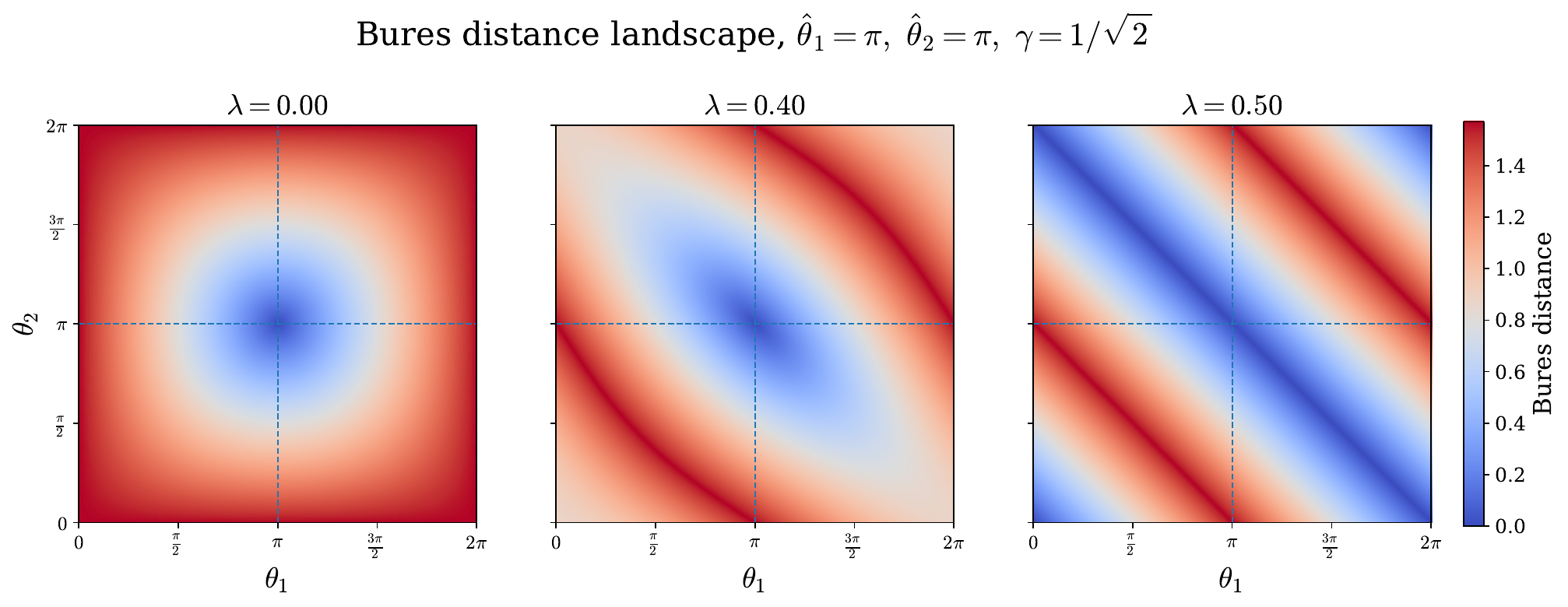}
    \caption{
    Images of Bures distance landscapes for $M=2$ mode states in the parameter space $\Theta$, centered at $\hat\theta_1 = \hat\theta_2 = \pi$, for the family of states of Eq.~\eqref{def:family} at $\gamma = 1/\sqrt{2}$ and distinct amounts of entanglement parametrized by $\lambda$. 
    From left to right: separable state ($\lambda = 0$), entangled state ($\lambda = 0.40$) and Bell state ($\lambda = 0.50$). 
    As the entanglement grows, the region of $\epsilon$-indistinguishable states deforms anisotropically along the QFI eigenvectors $w^{\pm}$: isotropic for the separable state, stretched for the entangled one, and collapsed into diagonal stripes on the parameter torus for the Bell state, where the direction $w^-$ associated with $\theta_1 - \theta_2$ becomes unresolvable and hence private while $w^+$ stays accessible.
}
    \label{fig:blobs}
\end{figure*}

Fig.~\ref{fig:blobs} illustrates, for $M=2$ modes, how the $\epsilon$-balls of the state space $\mathcal{S}$ defined above are displaced from a reference state centered at $\hat\theta_1,\hat\theta_2 = \pi$, for $\gamma = 1/\sqrt{2}$ and three distinct values of $\lambda$. 
For the sake of visualization, here privacy corresponds to the area of a single $\epsilon$-ball, while the definition given in Eq.~\eqref{def:privacy} is global and given in terms of average over the images of $\epsilon$-balls.
For separable states (left panel), the image of the $\epsilon$-ball in parameter space is almost symmetric about $\hat\theta_i$. 
Since states separated by a distance smaller than $\epsilon$ are assumed to be indistinguishable, these states approximately form a circle around the reference state. 
The whole parameter space can be tiled isotropically by such circles; that is, the states can be resolved up to a fixed accuracy in every direction. 
The number of $\epsilon$-balls required to cover the space scales as the inverse of the ball area, i.e. as $\epsilon^{-2}$.
When entanglement is introduced at $\lambda = 0.4$ (central panel), the image of the $\epsilon$-ball deforms anisotropically along the direction $w^+$ of the QFI eigenvectors, thus enabling an unbalanced estimation accuracy along the directions $w^{\pm}$. 
Nevertheless, for small $\epsilon$ the number of distinguishable regions scales as in the separable case.
Finally, for Bell states (right panel), the degeneracy of the QFI turns the covering balls into stripes that follow $w^+$ on the parameter space $\Theta$, spanning its full extent. 
The estimation thereby becomes effectively one-dimensional: the direction $w^-$, associated with $\theta_1 - \theta_2$, becomes unresolvable and hence private, while $w^+$ remains accessible. 
In this case the number of $\epsilon$-balls needed to cover the space scales as $\epsilon^{-1}$.

\begin{figure*}
    \centering

   \subfloat[$M = 2$ modes.\label{fig:privacy_and_accessibility_M=2}]{
    \includegraphics[width=\linewidth]{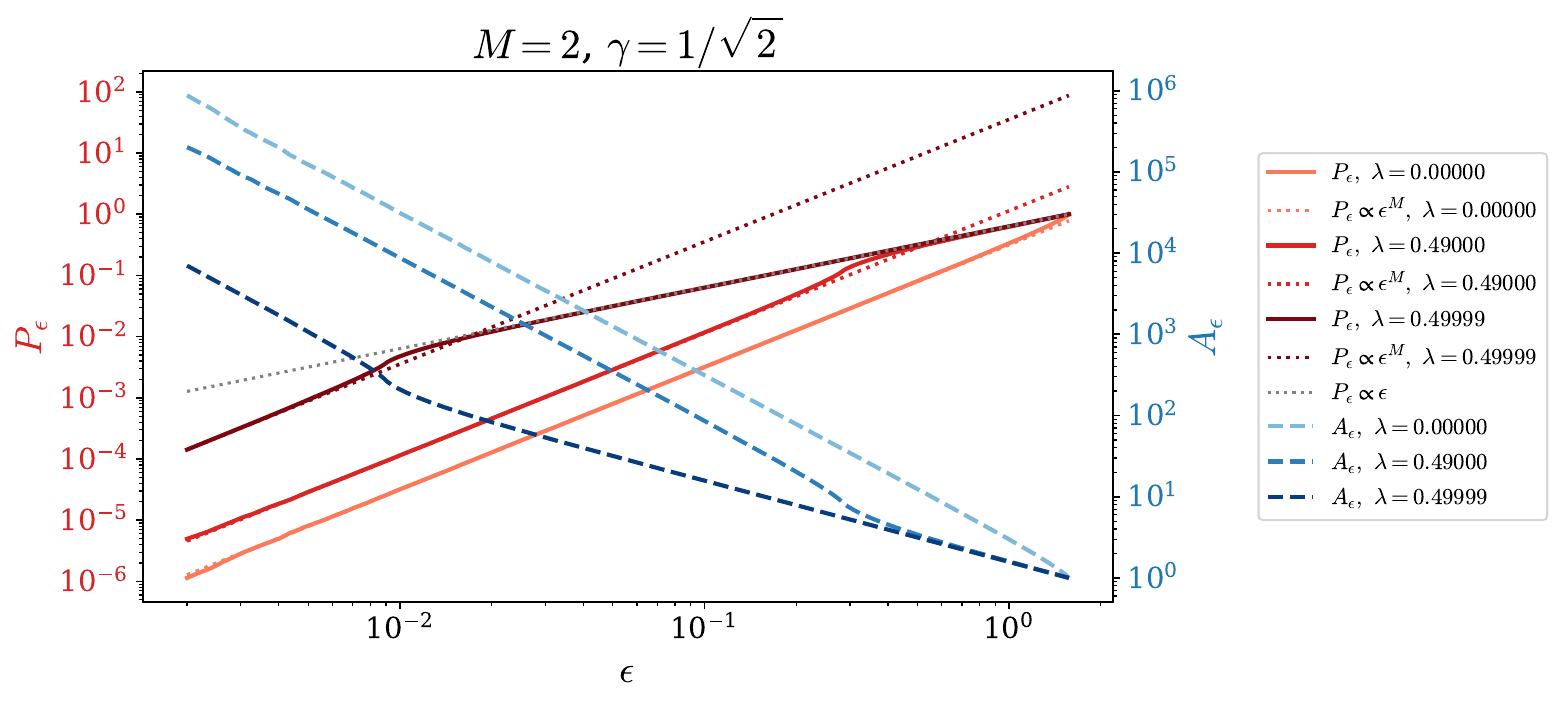}
}
\vspace{0.5cm}
\subfloat[$M = 4$ modes.\label{fig:privacy_and_accessibility_M=4}]{
    \includegraphics[width=\linewidth]{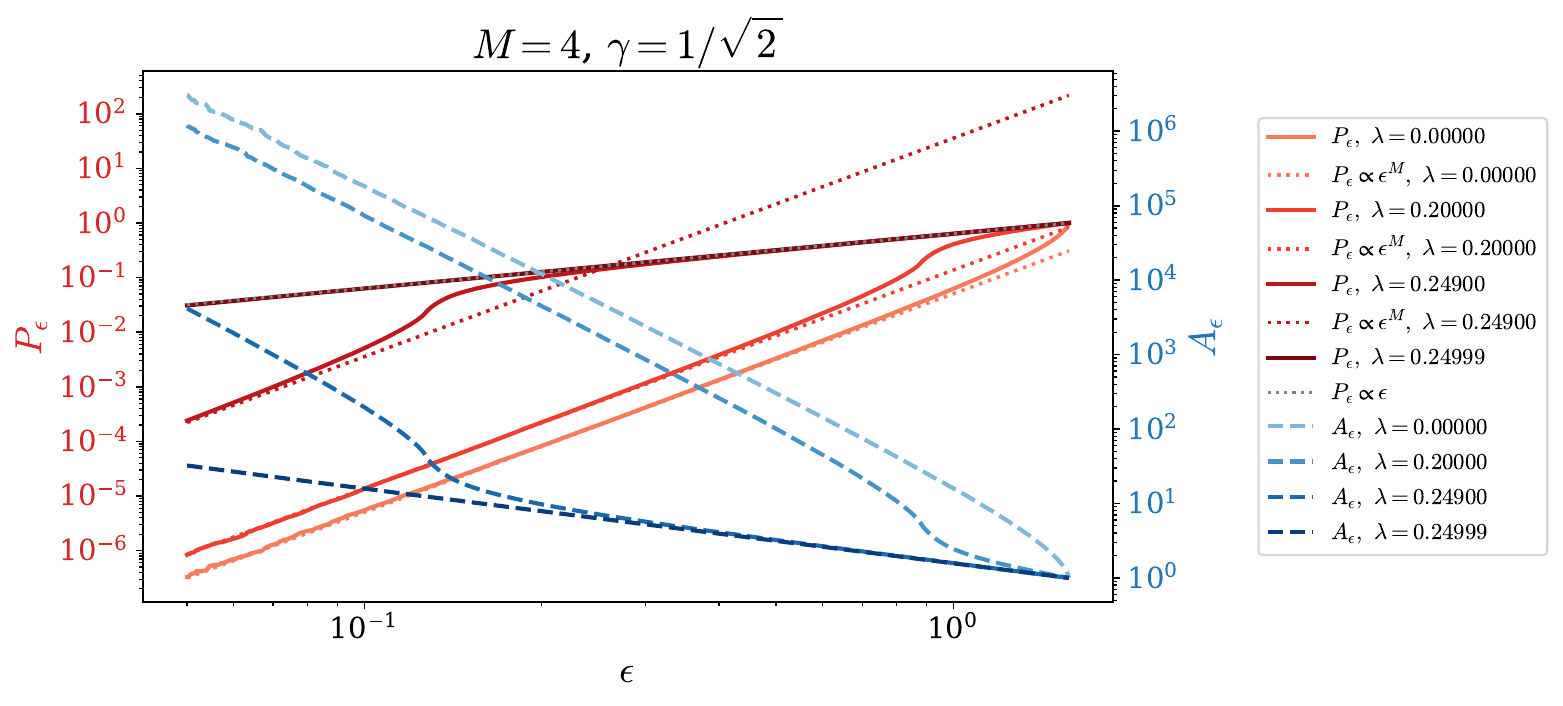}
}

\caption{
    $\epsilon$-privacy (Eq.~\eqref{def:privacy}, red solid lines) and $\epsilon$-accessibility (Eq.~\eqref{def:accessibility}, blue dashed lines) for the states of Eq.~\eqref{def:family} at $\gamma = 1/\sqrt{2}$, shown on a logarithmic scale as functions of $\epsilon$ for distinct values of $\lambda$, with (a) $M = 2$ and (b) $M = 4$ modes. 
    Privacy increases with both $\epsilon$ and $\lambda$, whereas accessibility decreases. 
    The reference curves proportional to $\epsilon^{M}$ (black dotted) and $\epsilon$ (grey dotted) indicate the limiting scalings: GHZ states attain maximal privacy and follow the $\epsilon$ scaling expected when states are resolved along a single direction, while separable states scale as $\epsilon^{M}$, consistent with an isotropic resolution. See Appendix \ref{appendix:epsilon-vall_volume} for the analytical detail.
    The kinks mark changes in the scaling law, which arise from the finite size of $\Theta$ as the uncertainty saturates the extent of the space along certain directions.
}
\label{fig:privacy_and_accessibility}
\end{figure*}

Fig.~\ref{fig:privacy_and_accessibility} shows, on a logarithmic scale, the $\epsilon$-privacy computed from Eq.~\eqref{def:privacy} and the $\epsilon$-accessibility from Eq.~\eqref{def:accessibility} for the states \eqref{def:family} at $\gamma = 1/\sqrt{2}$, as functions of $\epsilon$ for distinct values of $\lambda$. 
Privacy (red solid lines) increases with both $\epsilon$ and $\lambda$, whereas accessibility (blue dashed lines) decreases. 
To clarify their scaling laws, the reference curves proportional to $\epsilon^{M}$ (black dotted line) and $\epsilon$ (grey dotted line) are also displayed.
GHZ states attain maximal privacy at fixed $\epsilon$, confirming the $\epsilon$ scaling expected when states are resolved along a single direction only. 
Privacy of separable states, by contrast, scale as $\epsilon^{M}$, consistent with an isotropic resolution.
The changes in scaling, which appear as kinks in the plots, are a feature of the parameter space and of the encoding, originating in particular from the finite size of $\Theta$. 
As $\epsilon$ grows, the uncertainty saturates the extent of the space along certain directions, and the character of the scaling law changes accordingly.
The kink approaches smaller values of $\epsilon$ as the states approaches the ideal GHZ state limit. The detailed analysis of it is included in Appendix \ref{appendix:epsilon-vall_volume}, where the small $\epsilon$ asymptotic behavior denoted by dotted lines in Fig.~ \ref{fig:privacy_and_accessibility} is computed. 
This allows to write the analytical formula of privacy for states given by \eqref{def:family_maintext} when $\epsilon$ is sufficiently small:
\begin{equation}
\label{eq:privacy_separable}
    P_{\epsilon}\simeq\dfrac{\omega_M}{\pi^M\sqrt{\det g}}\epsilon^M\quad \epsilon\ll 1, \, q<1,
\end{equation}
where $q:=2^M\lambda^2$. 
Here,
\begin{equation}
    \omega_M=\dfrac{\pi^{M/2}}{\Gamma(M/2+1)}
\end{equation}
is the volume of the $M$-dimensional unit sphere and $\Gamma(\cdot)$ is the Euler function. 
For perfect GHZ states,
\begin{equation}
    P_{\epsilon}=\dfrac{2\epsilon}{\pi}\quad (q=c=1),
\end{equation}
where $c:=4\gamma^2(1-\gamma^2)$.

\section{Conclusions and Outlook}
    \label{conclusions}
    In this work, a notion of effective privacy and accessibility for quantum metrology has been introduced, where ``effective'' is understood as holding up to a finite accuracy in state discrimination. 
    This accuracy is quantified by a parameter $\epsilon$, which can be linked to the number of repetitions of the metrological protocol through a Chernoff-type argument and thereby related to the Bures distance. 
    The resulting figures of merit are information-geometric in nature: they are obtained by quantifying volumes in the parameter space, whose geometry is induced by the underlying states in Hilbert space. 
    This construction extends previous definitions of privacy, naturally accommodates imperfect implementations, and applies to more elaborate estimation scenarios, such as the encoding of nonlinear functions. 
    Through the representative example of extended-GHZ states, the behavior of these quantities has been illustrated, recovering the common intuition about privacy in quantum networks, and explicit scaling laws for the privacy have been obtained.
    Several questions remain open. A natural direction is to bridge the present geometric notion of privacy with its information-theoretic counterpart; in particular, the Rényi entropy provides an appealing link between geometry and information theory that seems worth exploring. 
    A further line concerns the role of state-preparation imperfections and of the distribution stage, the latter having been assumed ideal throughout this work.

\section{Acknowledgments}
We thank Yoshihiro Ueda for valuable discussions.
LB and DB acknowledge support from the European Union ERC StG, QOMUNE, 101077917. SY, WR, and MT acknowledge support from JST Moonshot R\&D,
Grant No. JPMJMS226C and Grant No. JPMJMS2061,
JST ASPIRE, Grant No. JPMJAP2427, JST COINEXT Grant No. JPMJPF2221, and  JST
CRONOS, Grant No. JPMJCS24N6.

\appendix

\section{Selected concepts of differential geometry}
\label{appendix:dif_geo}

This appendix offers a gentle introduction to the core differential-geometric
concepts used throughout the paper. 
For rigorous definitions and proofs the reader is referred to, e.g., \cite{fecko2006differential}.

An $M$-dimensional differentiable manifold $\mathcal{S}$ is a Hausdorff space
equipped with an atlas: a collection of homeomorphisms $\phi$
(charts) mapping open subsets of $\mathcal{S}$ onto open subsets of
$\mathbb{R}^M$, with smooth transition maps. 
Charts allow one to extend the
ordinary calculus of $\mathbb{R}^M$ onto $\mathcal{S}$. 
A real function
$f:\mathcal{S}\to\mathbb{R}$ 
is handled through its coordinate representative
\begin{equation}
    f\circ\phi^{-1} 
    \,:\, 
    \phi(U) \subseteq \mathbb{R}^M \longrightarrow \mathbb{R}.
\end{equation}
Together with functions, one introduces local objects, defined in a
neighborhood of a point, such as tangent vectors, forms and, more generally,
tensors. 
A tangent vector $v$ at $x \in \mathcal{S}$ is a derivation, that is, a linear map
sending a smooth function to a directional derivative and obeying the Leibniz
rule 
$v(fg)=v(f)\,g(x)+f(x)\,v(g)$,
\begin{equation}
    \begin{split}
    v :\ & C^\infty(\mathcal{S}) \longrightarrow \mathbb{R}\\
    & f \longmapsto v(f) = \sum_{i=1}^{M} v^{i}\partial_{x_i} f .
    \end{split}
\end{equation}
The basis vectors $\partial_{x_i}$ are referred as elements of a coordinate basis.
Being linear, the tangent vectors at $x$ form a vector space, the
tangent space $T_x(\mathcal{S})$.

A one-form is a linear map that takes in input a tangent vector and outputs a scalar,
\begin{equation}
    \begin{split}
        \omega :\ & T_x(\mathcal{S}) \longrightarrow \mathbb{R}\\
        & v \longmapsto \omega(v) = \sum_{i=1}^{M} \omega_i v^{i}.
    \end{split}
\end{equation}
One-forms form a vector space too, called $T^*_x(\mathcal{S})$, which turns out to be dual to $T_x(\mathcal{S})$,
with coordinate basis $\{dx^i\}$ fixed by $dx^i(\partial_{x_j})=\delta^i_j$.
Tensors are built by taking tensor products of vectors and one-forms. 
In particular, tensoring $p$ one-forms and antisymmetrizing yields a $p$-form, a totally antisymmetric multi-linear map
\begin{equation}
    \begin{split}
    \tilde{\omega} :\ & \underbrace{T_x(\mathcal{S})\otimes\cdots\otimes
        T_x(\mathcal{S})}_{p\text{ times}} \longrightarrow \mathbb{R}\\
    & (v_1,\dots,v_p)\longmapsto
        \tilde{\omega}(v_1,\dots,v_p)
        =
        \sum_{i_1<\dots<i_p}\tilde{\omega}_{i_1 \dots i_p}
        \tilde v^{i_1\dots i_p}
    \end{split}
\end{equation}
with 
$
v^{i_1\dots i_p} 
= 
\det{ 
    \left[ 
        v_b^{\,i_a}
    \right]_{a,b=1}^{p}
    }.
$
The antisymmetry makes $p$-forms the natural infinitesimal volume elements for
the integration of functions on the manifold. 
In particular, on an
$M$-dimensional manifold the space of $M$-forms is one-dimensional: there exists
a unique (up to rescaling) form $\tilde\omega$, which fixes an integration
measure $\mu$ for $\mathcal{S}$ through $d\mu=\tilde\omega$.

Furthermore, the duality between vectors and one-forms can be made canonical by endowing the
manifold with a metric tensor, which is a symmetric bilinear map eating two
vectors and returning a scalar,
\begin{equation}
    \begin{split}
        g :\ & T_x(\mathcal{S})\otimes T_x(\mathcal{S}) \longrightarrow \mathbb{R}\\
        & (v_1,v_2)\longmapsto g(v_1,v_2)
        =
        \sum_{i,j=1}^{M} g_{ij}v_1^{i}v_2^{j}.
    \end{split}
\end{equation}
Whenever it is non-degenerate, $g$ establishes a natural scalar product on each
tangent space and, in turn, singles out a canonical volume form,
\begin{equation}
    \tilde\omega 
    = 
    \sqrt{\big|\det[g]\big|} 
    dx^1\dots dx^M ,
\end{equation}
so that the associated infinitesimal measure reads $d\mu=\sqrt{|\det[g]|}d^{M}x$.

One can now relate objects living on different manifolds. 
Let $\mathcal{S}$ and $\mathcal{R}$ be two manifolds and
\begin{equation}
    f:\mathcal{S}\longrightarrow\mathcal{R}
\end{equation}
a smooth, not necessarily invertible, map. 
Consider a function $\psi:\mathcal{R}\to\mathbb{R}$. 
To evaluate it from the point of view of $\mathcal{S}$, the natural choice ---the only one that does not require $f^{-1}$--- is to pre-compose with $f$. 
This defines the pull-back of a function,
\begin{equation}
    f^{*}\psi := \psi\circ f \ :\ \mathcal{S}\longrightarrow\mathbb{R}.
\end{equation}
The name reflects that $\psi$, originally living on $\mathcal{R}$, is carried
``back" to $\mathcal{S}$.
Tangent vectors, by contrast, are transported forward: one simply drags the base point along through $f$. 
The resulting push-forward $f_*v\in T_{f(x)}(\mathcal{R})$ of a vector $v\in T_x(\mathcal{S})$ is defined by its action on functions of the target,
\begin{equation}
    \begin{split}
        f_{*} :\ & T_x(\mathcal{S}) \longrightarrow T_{f(x)}(\mathcal{R})\\
        & v \longmapsto f_*v = v[f^{*}\psi],
    \end{split}
\end{equation}
for every smooth $\psi$ on $\mathcal{R}$.
Since one-forms are dual to vectors, they are naturally pulled-back by $f$; and because the metric and the $p$-forms are obtained by tensoring one forms, they are pulled back as well. 
Explicitly, $(f^{*}\omega)(v)=\omega(f_*v)$ and $(f^{*}g)(v_1,v_2)=g(f_*v_1,f_*v_2)$.
Finally, let $\mu$ be a measure on $\mathcal{S}$, i.e. a map assigning to each measurable subset a number in $[0,\infty)$. For a measurable $R\subseteq\mathcal{R}$, the push-forward measure is defined as
\begin{equation}
    (f_{*}\mu)(R) := \mu\big(f^{-1}(R)\big),
\end{equation}
where $f^{-1}(R)\subseteq\mathcal{S}$ is the pre-image of $R$ under $f$. 
Measures therefore push-forward rather than pull-back, which is precisely what is needed to compensate the pull-back of the integrand, so that the change of variables
\begin{equation}
    \int_{\mathcal{R}} d(f_{*}\mu) \psi
    = 
    \int_{\mathcal{S}}
        d\mu
        (\psi\circ f) 
    = 
    \int_{\mathcal{S}} 
    d\mu 
    (f^{*}\psi)
\end{equation}
holds. 

\section{Properties of privacy measure}
\label{appendix:distinguishability}

The Fuchs-van de Graff inequality \cite{fuchs1999cryptographic} between the fidelity and the trace distance
\begin{equation}
\frac{1}{2}\|\rho-\rho'\|_{\mathrm{tr}}\leq \sqrt{1-F(\rho,\rho')}
\end{equation}
implies that
\begin{equation}
\frac{1}{2}\|\rho-\rho'\|_{\mathrm{tr}}\leq D(\rho,\rho').
\end{equation}
Therefore, if two states are closer than $\epsilon$ with respect to the Bures distance, so they are in terms of the trace distance. 
Moreover, the Helstrom bound \cite{helstrom1969quantum} states that the minimum error probability of distinguishability between two quantum states $\rho$ and $\rho'$ with balanced prior probabilities is given by
\begin{equation}
P_{e, \mathrm{min}}=\frac{1}{2}\left(1-\frac{1}{2}\|\rho-\rho'\|_{\mathrm{tr}}\right).
\end{equation}
Thus, $\epsilon$ bounded Bures distance implies
\begin{equation}
P_{e,min}\geq \frac{1}{2}-\frac{1}{2}\epsilon,
\end{equation}
which justifies treating all states in a ball
\begin{equation}
B_\epsilon(\rho)=\{\rho':D(\rho,\rho')\leq \epsilon\}
\label{def:epsilonball}
\end{equation}
as indistinguishable at the level of $\epsilon$. 

Furthermore, any quantum information processing including noise can be described by a completely positive and trace preserving (CPTP) map. 
Privacy defined from \eqref{def:epsilonvolume} inherits the contractivity property from the  Bures distance \cite{nielsen2010quantum}.
Intuitively, the privacy increases because more states become $\epsilon$-indistinguishable such operations.
This monotonicity implies that the $\epsilon$-privacy defined above cannot decrease under interception and processing of the quantum states assuming no extra information on the side of adversary which should be guaranteed by detailed analysis of the state preparation and sharing protocol. 
Also, since the fidelity, the Bures distance, and the QFI \cite{rezakhani2019continuity} are continuous under infinitesimal CPTP maps, so is the measure of the $\epsilon$-ball in the set of quantum states $\rho_{\theta}$ and their average $\epsilon$-privacy. 
The continuity, in turn, guarantees stability of privacy under small imperfections in state preparation.

\section{Privacy for imperfect GHZ states}
    \label{appendixA:application_details}
    Consider the family of states 
    \begin{equation}
    \label{def:family}
            \ket{\Psi_{\gamma, \lambda,\theta}} 
            = 
            A_{\lambda, \gamma}
            \ket{0}^{\otimes M}
            + 
            B_{\lambda,\gamma,\theta}
            \ket{1}^{\otimes M}
            +
            \sum_{x\in \mathcal{P}_M}
                C_{x, \lambda,\gamma,\theta}
                \ket{x},
    \end{equation}
    where $\mathcal{P}_M = \{0,1\}^{M}\setminus\{0^M,1^M\}$ and
    \small{
    \begin{equation}
        \begin{cases}
            A_{\lambda, \gamma} = \sqrt{2^M\lambda^2\,\gamma^2+(1-2^M\lambda^2)\,\gamma^{2M}} \\
            B_{\lambda,\gamma,\theta} = \sqrt{2^M\lambda^2(1-\gamma^2)+(1-2^M\lambda^2)(1-\gamma^2)^M}
                e^{i\sum_{j=1}^M\theta_j} \\
            C_{x,\lambda, \gamma,\theta} = \sqrt{1-2^M\lambda^2}\,\gamma^{M-|x|}(1-\gamma^2)^{|x|/2}
                e^{i\sum_{j=1}^M x_j\theta_j}\\
        \end{cases}
    \end{equation}
    }
    and $|x| = \sum_j x_j$ is the Hamming weight.
    At $\lambda=0$ one gets the separable state 
    $
    \ket{+_{\theta}}^{\otimes M}
    =
    \bigotimes_{j=1}^M
        (
            \gamma\ket{0}+\sqrt{1-\gamma^2}e^{i\theta_j}\ket{1}
        )
    $,
    while at $\lambda=2^{-M/2}$ one obtains the GHZ-like state
    $
    \ket{\Psi_{\gamma,\theta}}
    = 
    \gamma\ket{0}^{\otimes M}
    +
    \sqrt{1-\gamma^2}e^{i\sum_{i=1}^N\theta_i}\ket{1}^{\otimes M},
    $
    with the usual GHZ state recovered for $\gamma=1/\sqrt{2}$.
    
    The QFI matrix for the state in \eqref{def:family} is 
    \begin{equation}\label{def:qfi_matrix_family}
        q_{ij} 
        = 
        4\gamma^2(1-\gamma^2)
            \left[
                (1-2^M\lambda^2)
                \mathrm{id}_M 
                + 
                2^M\lambda^2
                \mathrm{J}_M
            \right],
    \end{equation}
    with $\mathrm{J}_M = \mathbf{1}\mathbf{1}^{\T}$ and $\mathbf{1}^{\T}$ the $M$-dimensional vector $\mathbf{1}^{\T} = (1,\dots,1)$.
    The spectrum of the QFI has an eigenvalue
    $\mu^+ = 4\gamma^2(1-\gamma^2)[1+(M-1)2^M\lambda^2]$, with multiplicity one associated to the eigenvector $w^+ = (1/\sqrt{M})(1,1,...,1)^T$ and an eigenvalue $\mu^- = 4\gamma^2(1-\gamma^2)[1-2^M\lambda^2]$ with multiplicity $M-1$, associated to the eigenspace spanned by 
    $w^k = e_1 - e_{k+1}$, with $e_k = (0,\dots,1,\dots,0)$, the non-null element in position $k$.
    
    After classical communication, the central user of the network can then construct the state based on a set of estimators $\{\hat\theta_i\}_{i=1}^M$ as $\ket{\Psi_{\gamma, \lambda,\hat\theta}}$. 
    They can estimate the distance between this state and the $\theta$-dependent state by means of the Bures distance and the related fidelity, making them functions of the parameters
    
    \begin{equation}
        \label{example:sqrt_fidelity}
        \begin{split}
                F(\Psi_{\gamma, \lambda,\theta},\Psi_{\gamma, \lambda,\hat\theta})
                ={}&
                \left|
                    2^M\lambda^2
                    \left[
                        \gamma^2 + (1-\gamma^2)e^{i\sum_{k=1}^M\phi_k}
                    \right]+\right.\\
                   &\left.+(1-2^M\lambda^2)
                   \prod_{k=1}^M 
                        \left[
                        \gamma^2 
                        + 
                        (1-\gamma^2)
                        e^{ 
                            i\phi_k
                        }
                        \right]
                \right|^2
        \end{split}
    \end{equation}
    with $\phi_k = \theta_k-\hat\theta_k$.
    
    Furthermore, one may derive the necessary ingredients for the computation of directional privacy.
    The eigenvector $w^+$ is orthogonal to a family of hypersurfaces $\Sigma^+: \sum_{i=1}^M \theta_i = c$. 
    This is a family of $p= M-1$ dimensional hyperplanes conveniently parametrized by
    \begin{equation}
        \begin{cases}
            \theta_1 = \alpha_1\\ 
            \theta_2 = \alpha_2 \\ 
            \vdots\\ 
            \theta_M = c - \sum_{i=1}^{M-1} \alpha_i
        \end{cases}.
    \end{equation}
    The Jacobian parametrizing the change of coordinates on this submanifold is given by the $M \times (M-1)$ matrix 
    \begin{equation}
        J^+ 
        = 
        \begin{bmatrix}
        1 & 0 &\cdots &0 \\ 
        0 & 1 & \cdots & 0 \\ 
        \vdots & \vdots & \cdots & \vdots \\ 
        -1 & -1 &\cdots &-1
        \end{bmatrix}
    \end{equation}
    and  the reduced QFI on $\Sigma^+$ is thus 
    
    \begin{equation}
        h^+_{ab} 
        = 
        4\gamma^2(1-\gamma^2)
        \begin{bmatrix}
            \dfrac{1-2^M\lambda^2}{2} & \dfrac{1-2^M\lambda^2}{4} & \cdots & \dfrac{1-2^M\lambda^2}{4}\\
            \dfrac{1-2^M\lambda^2}{4} & \dfrac{1-2^M\lambda^2}{2} & \cdots & \dfrac{1-2^M\lambda^2}{4}\\
             \vdots & \vdots & \ddots & \vdots \\
            \dfrac{1-2^M\lambda^2}{4} & \dfrac{1-2^M\lambda^2}{4} & \cdots & \dfrac{1-2^M\lambda^2}{2}
        \end{bmatrix},
    \end{equation}
    
    yielding a volume element of
    
    \begin{equation}
        \sqrt{|\det h^+|} 
        =
        \left[
            4\gamma^2(1-\gamma^2)
        \right]^{(M-1)/2}
        \sqrt{\frac{M(1-2^M\lambda^2)^{M-1}}{4^{M-1}}}
    \end{equation}
    
    The eigenvectors $w^k$ are each one orthogonal to a family of hypersurfaces $\Sigma^k: \theta_1-\theta_{k+1} = c'$ parametrized by 
    \begin{equation}
        \begin{cases}
        \theta_j = \beta_j \; \forall \; j \neq k+1 \\
        \theta_{k+1} = \beta_1 - c'
        \end{cases}.
    \end{equation}
    These coordinates give a Jacobian transformation 
    \begin{equation}
        J^- 
        = 
        \begin{bmatrix}
            1 & 0 & \cdots & 0 \\ 0 & 1 & \cdots & 0 \\ \vdots  & \vdots & \cdots & \vdots \\ 1 & 0 & \cdots & 0 \\ \vdots & \vdots &\cdots & \vdots \\ 0 & 0 & \cdots & 1
        \end{bmatrix},
    \end{equation}
    which yields the reduced QFI on the submanifold $\Sigma^k$:
    
    \begin{equation}
        h^-_{ab} 
        = 
        4\gamma^2(1-\gamma^2)
        \begin{bmatrix}
            \dfrac{1+2^M\lambda^2}{2} & 2^{M-1}\lambda^2 & \cdots & 2^{M-1}\lambda^2 \\
            2^{M-1}\lambda^2 & \dfrac{1}{4} & \cdots & 2^{M-2}\lambda^2 \\
            \vdots & \vdots & \ddots & \vdots \\
            2^{M-1}\lambda^2 & 2^{M-2}\lambda^2 & \cdots & \dfrac{1}{4}
        \end{bmatrix}
    \end{equation}
    whose volume element is

    \begin{equation}
        \begin{split}
            \sqrt{
                |\det h^-|
                }
            =&{}
            \Big(
                \left[
                    4\gamma^2(1-\gamma^2)
                \right]^{M-1}\\
            &\times\frac{
                (1-2^M\lambda^2)^{M-2}}
                {2^{2M-3}}\\
            &\times\left[
                1+(M-1)2^M\lambda^2
                \right]
            \Big)^{1/2}
\end{split}
\end{equation}

\section{Volumes of $\epsilon$-ball}
\label{appendix:epsilon-vall_volume}

    When considering the family of states given in \eqref{def:family}, the privacy in Eq.~\eqref{def:privacy} is given by
    \begin{equation}
        \begin{split}
            P_{\epsilon} &= \dfrac{1}{(2\pi)^M}\int_{[0,2\pi)^M} d^M\phi\,H[\epsilon-D_B(\phi)]\\
            &= \mathrm{Pr}_{\phi\sim\mathrm{Unif[[0,2\pi)^M]}}[D_B(\phi)\leq\epsilon],
        \end{split}
    \end{equation}
    where $\phi :=\theta-\hat{\theta} \mod 2\pi$. 
    Translational invariance of $\Theta$ combined with \emph{i.i.d.} parameters make the variable $\phi$ obey the uniform distribution $\mathrm{Unif[[0,2\pi)^M]}$ over $[0,2\pi)^M$. 
    \begin{equation}
        \theta, \hat{\theta}\overset{\mathrm{i.i.d.}}{\sim} \mathrm{Unif}[[0,2\pi)^M] \implies \phi\sim \mathrm{Unif}[[0,2\pi)^M].
    \end{equation}
    One can then compactly rewrite
    \begin{align}
        \det g = c^M(1-q)^{M-1}[1+(M-1)q],
    \end{align}
    where $c:= 4\gamma^2(1-\gamma^2), q:=2^M\lambda^2$.
    At $q=0$ (that is, $\lambda=0$) one gets the separable state, while at $q=1$ (corresponding to $\lambda=2^{-M/2}$) one obtains a GHZ-like state, with the usual GHZ state recovered for $c=1$, standing for $\gamma=1/\sqrt{2}$.
    
    First, consider $q=1$. 
    By using \eqref{example:sqrt_fidelity}, 
    \begin{equation}
        D_B(\phi)=\arccos \sqrt{1-c\sin^2\dfrac{S}{2}},
    \end{equation}
    where $S=\sum_{j=1}^M\phi_j \mod{2\pi}$. 
    By a simple calculation,
    \begin{equation}
        \begin{split}
            &\arccos \sqrt{1-c\sin^2\dfrac{S}{2}}\leq\epsilon \\
            \iff &
            \begin{cases}
                S \in [0, 2\pi) &(c=0\, \lor \,\sin^2\epsilon\geq c), \\
                S\in [0,2\alpha] \cup [2\pi-2\alpha, 2\pi) & (0<c \,\land \, \sin^2\epsilon < c),
            \end{cases}
        \end{split}
    \end{equation}
    where $\alpha=\arcsin \frac{\sin\epsilon}{c}$ and
    \begin{equation}
        \phi\sim\mathrm{Unif}[[0,2\pi)^M] \implies S\sim\mathrm{Unif}[[0,2\pi)].
    \end{equation}
    Thus,  
    \begin{equation}\label{eq:privacy_GHZ}
        \begin{split}
            P_{\epsilon}&=\mathrm{Pr}\left[\arccos \sqrt{1-c\sin^2\dfrac{S}{2}}\leq\epsilon\right]\\
            &=
            \begin{cases}
                1 &(c=0\, \lor \,\sin^2\epsilon\geq c), \\
                \dfrac{2\alpha}{\pi} & (0<c \,\land \, \sin^2\epsilon < c),
            \end{cases}
             \quad (q=1).
        \end{split}
    \end{equation}
    For the usual GHZ state, $c=1$, the privacy reduces to
    \begin{equation}\label{eq:privacy_perfect_GHZ}
        P_{\epsilon}=\dfrac{2\epsilon}{\pi}\quad (q=c=1).
    \end{equation}
    
    Next consider the separable case ($q<1$) under which the QFI matrix $g$ is non-degenerate.
    It is known \cite{liu2020quantum} that
    \begin{equation}
        D_B(\phi)^2 = \dfrac{1}{4}\phi^{\T}g\phi + o(\|\phi\|^2).
    \end{equation}
    Therefore, when $\epsilon\ll 1$, 
    \begin{equation}
        \label{eq:M-dimensional_ellipse}
        D_B(\phi)\leq\epsilon \iff \phi^{\T}g\phi\leq 4\epsilon^2
    \end{equation}
    and the right-hand side represents an $M$-dimensional ellipse whose volume is given by
    \begin{equation}
        \mathrm{Vol}\{\phi : \phi^{\T}g\phi\leq 4\epsilon^2\}=\dfrac{\omega_M(2\epsilon)^M}{\sqrt{\det g}}
    \end{equation}
    for $g$ non-degenerate. 
    Here,
    \begin{equation}
        \omega_M=\dfrac{\pi^{M/2}}{\Gamma(M/2+1)}
    \end{equation}
    is the volume of $M$-dimensional unit sphere and $\Gamma(\cdot)$ is the Euler function. 
    This leads to a closed-form expression for the privacy 
    \begin{equation}
        \label{eq:privacy_separable}
        P_{\epsilon}\simeq\dfrac{\omega_M}{\pi^M\sqrt{\det g}}\epsilon^M\quad (\epsilon\ll 1, q<1).
    \end{equation}
    Eqs.~\eqref{eq:privacy_perfect_GHZ} and \eqref{eq:privacy_separable} are plotted for $\gamma=1/\sqrt{2}$ and distinct values of $\lambda$ as dotted lines. 
    The analytical solutions coincides with the numerical results of the privacy (solid lines) shown in Fig.~ \ref{fig:privacy_and_accessibility} of the main text. 
    
    The numerical plots show that the scaling of privacy changes from $\epsilon^M$ to $\epsilon$, as $\epsilon\to\pi/2$ and $  q\to 1$. 
    This can be understood in terms of the QFI matrix \eqref{def:qfi_matrix_family}, which has an eigenvalue $\mu^+$ that corresponds to the eigenvector $w^+$ and $M-1$ eigenvalues $\mu^-$ associated to $M-1$ directions $v^k$, which can be chosen as 
    \begin{equation}
        v^k = \frac{1}{\sqrt{k(k+1)}}(1,\dots,1,-k,0,\dots,0),
    \end{equation}
    normalized vectors with the first $k-1$ components equal to one, the $k$-th component equal to $-k$ and the others equal to zero.
    With the decomposition 
    \begin{equation}
        \phi
        = u w^+ + \sum_{k=1}^{M-1}z_kv^k,
    \end{equation}
     the region $D_B(\phi)\leq\epsilon$ \eqref{eq:M-dimensional_ellipse} is expressed as  
    \begin{equation}
        \left(\dfrac{u}{R^{+}}\right)^2+\sum_{k=1}^{M-1}\left(\dfrac{z_k}{R^{-}}\right)^2\leq 1
    \end{equation}
    with $R^{+} = 2\epsilon/\sqrt{\mu^+}$ and $R^{-} = 2\epsilon/\sqrt{\mu^-}$.
    This equation describes an  $M$-dimensional ellipse in $(u, z_1, \dots, z_{M-1})$ space. 
    As $q\to1$, the radii in $z_i$-directions diverge: $R^{-}\overset{q\to 1}{\to}\infty$, whereas the radius in $u$ direction does not: $R^{+}\overset{q\to 1}{\to} \epsilon/\sqrt{cM}$. 
    Consequently, as $\epsilon$ increases, the ellipsoid first extends across the periodic parameter space $[0,2\pi)^M$ in $z_i$-directions. Beyond this crossover, only its extent along $u$-direction continues to grow with $\epsilon$. Therefore, the volume scaling changes effectively from $\epsilon^M$ to $\epsilon$. 
    The same happens when the value of $q$ is fixed while $\epsilon$ increases.

\bibliography{biblio}

@article{liu2020quantum,
  title={Quantum Fisher information matrix and multiparameter estimation},
  author={Liu, Jiang and Yuan, Haidong and Lu, Xiao-Ming and Wang, Xiaoguang},
  journal={Journal of Physics A: Mathematical and Theoretical},
  volume={53},
  number={2},
  pages={023001},
  year={2020},
  publisher={IOP Publishing},
  doi = {10.1088/1751-8121/ab5d4d}
}

@article{leonenko2010statistical,
  title={Statistical inference for the $\epsilon$-entropy and the quadratic R{\'e}nyi entropy},
  author={Leonenko, Nikolaj and Seleznjev, Oleg},
  journal={Journal of Multivariate Analysis},
  volume={101},
  number={9},
  pages={1981--1994},
  year={2010},
  publisher={Elsevier},
  doi = {10.1016/j.jmva.2010.05.009}
}

@article{luczak1997suboptimal,
  title={A suboptimal lossy data compression based on approximate pattern matching},
  author={Luczak, Tomasz and Szpankowski, Wojciech},
  journal={IEEE transactions on Information Theory},
  volume={43},
  number={5},
  pages={1439--1451},
  year={1997},
  publisher={IEEE},
  doi = {10.1109/18.623143}
}

@article{seleznjev2010random,
  title={Random databases with approximate record matching},
  author={Seleznjev, Oleg and Thalheim, Bernhard},
  journal={Methodology and Computing in Applied Probability},
  volume={12},
  number={1},
  pages={63--89},
  year={2010},
  publisher={Springer},
  doi = {10.1007/s11009-008-9092-4}
}

@article{zyczkowski2003renyi,
  title={R{\'e}nyi extrapolation of Shannon entropy},
  author={{\.Z}yczkowski, Karol},
  journal={Open Systems \& Information Dynamics},
  volume={10},
  number={3},
  pages={297--310},
  year={2003},
  publisher={Springer},
  doi = {
https://doi.org/10.1023/A:1025128024427}
}

@article{azuma2023quantum,
  title={Quantum repeaters: From quantum networks to the quantum internet},
  author={Azuma, Koji and Economou, Sophia E and Elkouss, David and Hilaire, Paul and Jiang, Liang and Lo, Hoi-Kwong and Tzitrin, Ilan},
  journal={Reviews of Modern Physics},
  volume={95},
  number={4},
  pages={045006},
  year={2023},
  publisher={APS},
  doi = {https://doi.org/10.1103/RevModPhys.95.045006}
}

@article{barral2025review,
  title={Review of distributed quantum computing: From single qpu to high performance quantum computing},
  author={Barral, David and Cardama, F Javier and D{\'\i}az-Camacho, Guillermo and Fa{\'\i}lde, Daniel and Llovo, Iago F and Mussa-Juane, Mariamo and V{\'a}zquez-P{\'e}rez, Jorge and Villasuso, Juan and Pi{\~n}eiro, C{\'e}sar and Costas, Natalia and others},
  journal={Computer Science Review},
  volume={57},
  pages={100747},
  year={2025},
  publisher={Elsevier},
  doi = {10.1016/j.cosrev.2025.100747}
}

@article{degen2017quantum,
  title={Quantum sensing},
  author={Degen, Christian L and Reinhard, Friedemann and Cappellaro, Paola},
  journal={Reviews of modern physics},
  volume={89},
  number={3},
  pages={035002},
  year={2017},
  publisher={APS},
  doi = {https://doi.org/10.1103/RevModPhys.89.035002}
}

@article{zhang2021distributed,
  title={Distributed quantum sensing},
  author={Zhang, Zheshen and Zhuang, Quntao},
  journal={Quantum Science \& Technology},
  volume={6},
  number={4},
  pages={043001},
  year={2021},
  publisher={IOP Publishing},
  doi = {10.1088/2058-9565/abd4c3}
}

@article{komar2014quantum,
  title={A quantum network of clocks},
  author={Komar, Peter and Kessler, Eric M and Bishof, Michael and Jiang, Liang and S{\o}rensen, Anders S and Ye, Jun and Lukin, Mikhail D},
  journal={Nature Physics},
  volume={10},
  number={8},
  pages={582--587},
  year={2014},
  publisher={Nature Publishing Group UK London},
  doi = {https://doi.org/10.1038/nphys3000}
}

@inproceedings{azahari2024review,
  title={Review of clock synchronization in quantum communications},
  author={Azahari, Nur Shahirah and Harun, Nur Ziadah and Chai Wen, Chuah and Ramli, Sofia Najwa and Ahmad Zukarnain, Zuriati},
  booktitle={Proceedings of the 2024 13th International Conference on Software and Computer Applications},
  pages={350--356},
  year={2024},
  doi = {https://doi.org/10.1145/3651781.3651834}
}

@article{hassani2025privacy,
  title={Privacy in networks of quantum sensors},
  author={Hassani, Majid and Scheiner, Santiago and Paris, Matteo GA and Markham, Damian},
  journal={Physical Review Letters},
  volume={134},
  number={3},
  pages={030802},
  year={2025},
  publisher={APS},
  doi = {https://doi.org/10.1103/PhysRevLett.134.030802}
}

@article{huang2019cryptographic,
  title={Cryptographic quantum metrology},
  author={Huang, Zixin and Macchiavello, Chiara and Maccone, Lorenzo},
  journal={Physical Review A},
  volume={99},
  number={2},
  pages={022314},
  year={2019},
  publisher={APS},
  doi = {https://doi.org/10.1103/PhysRevA.99.022314}
}

@article{shettell2022cryptographic,
  title={Cryptographic approach to quantum metrology},
  author={Shettell, Nathan and Kashefi, Elham and Markham, Damian},
  journal={Physical Review A},
  volume={105},
  number={1},
  pages={L010401},
  year={2022},
  publisher={APS},
  doi = {https://doi.org/10.1103/PhysRevA.105.L010401}
}

@article{moore2023secure,
  title={Secure quantum remote sensing without entanglement},
  author={Moore, Sean W and Dunningham, Jacob A},
  journal={AVS Quantum Science},
  volume={5},
  number={1},
  year={2023},
  publisher={AIP Publishing},
  doi = {https://doi.org/10.1116/5.0137260}
}

@article{shettell2022private,
  title={Private network parameter estimation with quantum sensors},
  author={Shettell, Nathan and Hassani, Majid and Markham, Damian},
  journal={arXiv preprint arXiv:2207.14450},
  year={2022},
  doi = {
https://doi.org/10.48550/arXiv.2207.14450}
}

@article{toth2014quantum,
  title={Quantum metrology from a quantum information science perspective},
  author={T{\'o}th, G{\'e}za and Apellaniz, Iagoba},
  journal={Journal of Physics A: Mathematical and Theoretical},
  volume={47},
  number={42},
  pages={424006},
  year={2014},
  publisher={IOP Publishing},
  doi = {10.1088/1751-8113/47/42/424006}
}

@article{yin2020experimental,
  title={Experimental demonstration of secure quantum remote sensing},
  author={Yin, Peng and Takeuchi, Yuki and Zhang, Wen-Hao and Yin, Zhen-Qiang and Matsuzaki, Yuichiro and Peng, Xing-Xiang and Xu, Xiao-Ye and Xu, Jin-Shi and Tang, Jian-Shun and Zhou, Zong-Quan and others},
  journal={Physical Review Applied},
  volume={14},
  number={1},
  pages={014065},
  year={2020},
  publisher={APS},
  doi = {https://doi.org/10.1103/PhysRevApplied.14.014065}
}

@article{ho2026quantum,
  title={Quantum-private distributed sensing},
  author={Ho, Joseph and Webb, Jonathan W and Brooks, Russell MJ and Grasselli, Federico and Gauger, Erik and Fedrizzi, Alessandro},
  journal={Journal of Physics: Photonics},
  volume={8},
  number={2},
  pages={025006},
  year={2026},
  publisher={IOP Publishing},
  doi = {10.1088/2515-7647/ae551a}
}

@article{de2025anonymous,
  title={Anonymous and private parameter estimation in networks of quantum sensors},
  author={de Jong, Jarn and Scheiner, Santiago and Solomons, Naomi R and Chaoui, Ziad and Markham, Damian and Pappa, Anna},
  journal={Physical Review Applied},
  volume={24},
  number={5},
  pages={054053},
  year={2025},
  publisher={APS},
  doi = {https://doi.org/10.1103/lbfk-cykl}
}

@article{moore2025secure,
  title={Secure quantum-enhanced measurements on a network of sensors},
  author={Moore, Sean William and Dunningham, Jacob A},
  journal={Physical Review A},
  volume={111},
  number={1},
  pages={012616},
  year={2025},
  publisher={APS},
  doi = {https://doi.org/10.1103/PhysRevA.111.012616}
}

@article{bugalho2025private,
  title={Private and robust states for distributed quantum sensing},
  author={Bugalho, Lu{\'\i}s and Hassani, Majid and Omar, Yasser and Markham, Damian},
  journal={Quantum},
  volume={9},
  pages={1596},
  year={2025},
  publisher={Verein zur F{\"o}rderung des Open Access Publizierens in den Quantenwissenschaften},
  doi = {https://doi.org/10.22331/q-2025-01-15-1596}
}

@article{alushi2026privacy,
  title={Privacy in distributed quantum sensing with Gaussian quantum networks},
  author={Alushi, Uesli and Di Candia, Roberto},
  journal={npj Quantum Information},
  year={2026},
  publisher={Nature Publishing Group UK London},
  doi = {https://doi.org/10.1038/s41534-026-01266-3}
}

@article{junior2025privacy,
  title={Privacy in continuous-variable distributed quantum sensing},
  author={Junior, A and Andersen, Anton L and Larsen, Benjamin Lundgren and Moore, Sean William and Markham, Damian and Takeoka, Masahiro and Brask, Jonatan Bohr and Andersen, Ulrik L},
  journal={arXiv preprint arXiv:2509.12338},
  year={2025},
  doi = {
https://doi.org/10.48550/arXiv.2509.12338}
}

@article{namkung2026universal,
  title={Universal Operational Privacy in Distributed Quantum Sensing},
  author={Namkung, Min and Kim, Dong-Hyun and Hong, Seongjin and Kim, Yong-Su and Lee, Su-Yong and Lim, Hyang-Tag},
  journal={arXiv preprint arXiv:2601.19206},
  year={2026},
  doi = {
https://doi.org/10.48550/arXiv.2601.19206}
}

@article{solomons2025composable,
  title={Composable privacy of networked quantum sensing},
  author={Solomons, Naomi R and Markham, Damian},
  journal={arXiv preprint arXiv:2510.06326},
  year={2025},
  doi = {
https://doi.org/10.48550/arXiv.2510.06326}
}

@book{bengtsson2017geometry,
  title={Geometry of quantum states: an introduction to quantum entanglement},
  author={Bengtsson, Ingemar and {\.Z}yczkowski, Karol},
  year={2017},
  publisher={Cambridge university press},
  doi = {https://doi.org/10.1017/CBO9780511535048}
}

@article{braunstein1994statistical,
  title={Statistical distance and the geometry of quantum states},
  author={Braunstein, Samuel L and Caves, Carlton M},
  journal={Physical Review Letters},
  volume={72},
  number={22},
  pages={3439},
  year={1994},
  publisher={APS},
  doi = {https://doi.org/10.1103/PhysRevLett.72.3439}
}

@article{brody2001geometric,
  title={Geometric quantum mechanics},
  author={Brody, Dorje C and Hughston, Lane P},
  journal={Journal of geometry and physics},
  volume={38},
  number={1},
  pages={19--53},
  year={2001},
  publisher={Elsevier},
  doi ={https://doi.org/10.1016/S0393-0440(00)00052-8}
}

@article{silva2001lectures,
  title={Lectures on symplectic geometry},
  author={Silva, Ana Cannas},
  journal={Lecture Notes in Mathematics},
  volume={1764},
  year={2001},
  publisher={Springer},
  doi = {https://doi.org/10.1007/978-3-540-45330-7}
}

@article{bures1969extension,
  title={An extension of Kakutani's theorem on infinite product measures to the tensor product of semifinite w*-algebras},
  author={Bures, Donald},
  journal={Transactions of the American Mathematical Society},
  volume={135},
  pages={199--212},
  year={1969},
  publisher={JSTOR},
  doi = {https://doi.org/10.2307/1995012}
}

@article{helstrom1969quantum,
  title={Quantum detection and estimation theory},
  author={Helstrom, Carl W},
  journal={Journal of statistical physics},
  volume={1},
  number={2},
  pages={231--252},
  year={1969},
  publisher={Springer},
  doi = {https://doi.org/10.1007/BF01007479}
}

@article{wootters1981statistical,
  title={Statistical distance and Hilbert space},
  author={Wootters, William K},
  journal={Physical Review D},
  volume={23},
  number={2},
  pages={357},
  year={1981},
  publisher={APS},
  doi = {https://doi.org/10.1103/PhysRevD.23.357}
}

@book{fecko2006differential,
  title={Differential geometry and Lie groups for physicists},
  author={Fecko, Mari{\'a}n},
  year={2006},
  publisher={Cambridge university press},
  doi = {https://doi.org/10.1017/CBO9780511755590}
}

@inproceedings{wei2024investigations,
  title={Investigations on collision entropy and its variant for machine condition monitoring},
  author={Wei, Lan and Wang, Dong and Wang, Yu and Yan, Ming},
  booktitle={Journal of Physics: Conference Series},
  volume={2853},
  number={1},
  pages={012069},
  year={2024},
  organization={IOP Publishing}
}

@inproceedings{renyi1961measures,
  title={On measures of entropy and information},
  author={R{\'e}nyi, Alfr{\'e}d},
  booktitle={Proceedings of the fourth Berkeley symposium on mathematical statistics and probability, volume 1: contributions to the theory of statistics},
  volume={4},
  pages={547--562},
  year={1961},
  organization={University of California Press}
}

@article{bosyk2012collision,
  title={Collision entropy and optimal uncertainty},
  author={Bosyk, Gustavo Mart{\i}n and Portesi, M and Plastino, A},
  journal={Physical Review A—Atomic, Molecular, and Optical Physics},
  volume={85},
  number={1},
  pages={012108},
  year={2012},
  publisher={APS}
}

@article{ribeiro2021entropy,
  title={The entropy universe},
  author={Ribeiro, Maria and Henriques, Teresa and Castro, Lu{\'\i}sa and Souto, Andr{\'e} and Antunes, Lu{\'\i}s and Costa-Santos, Cristina and Teixeira, Andreia},
  journal={Entropy},
  volume={23},
  number={2},
  pages={222},
  year={2021},
  publisher={MDPI}
}

@book{nielsen2010quantum,
  title={Quantum computation and quantum information},
  author={Nielsen, Michael A and Chuang, Isaac L},
  year={2010},
  publisher={Cambridge university press},
  doi={https://doi.org/10.1017/CBO9780511976667}
}

@article{rezakhani2019continuity,
  title={Continuity of the quantum Fisher information},
  author={Rezakhani, AT and Hassani, Majid and Alipour, Sahar},
  journal={Physical Review A},
  volume={100},
  number={3},
  pages={032317},
  year={2019},
  publisher={APS},
  doi = {https://doi.org/10.1103/PhysRevA.100.032317}
}

@article{federer1959curvature,
  title={Curvature measures},
  author={Federer, Herbert},
  journal={Transactions of the American Mathematical Society},
  volume={93},
  number={3},
  pages={418--491},
  year={1959},
  publisher={JSTOR},
  doi = {https://doi.org/10.2307/1993504}
}

@article{fuchs1999cryptographic,
  title={Cryptographic distinguishability measures for quantum-mechanical states},
  author={Fuchs, Christopher A and Van De Graaf, Jeroen},
  journal={IEEE transactions on information theory},
  volume={45},
  number={4},
  pages={1216--1227},
  year={1999},
  publisher={IEEE},
  doi={10.1109/18.761271}
}

@article{horodecki2009quantum,
  title={Quantum entanglement},
  author={Horodecki, Ryszard and Horodecki, Pawe{\l} and Horodecki, Micha{\l} and Horodecki, Karol},
  journal={Reviews of modern physics},
  volume={81},
  number={2},
  pages={865--942},
  year={2009},
  publisher={APS},
  doi = {https://doi.org/10.1103/RevModPhys.81.865}
}

@article{shor1999polynomial,
  title={Polynomial-time algorithms for prime factorization and discrete logarithms on a quantum computer},
  author={Shor, Peter W},
  journal={SIAM review},
  volume={41},
  number={2},
  pages={303--332},
  year={1999},
  publisher={SIAM},
  doi = {https://doi.org/10.1137/S0036144598347011}
}

@article{ekert1991quantum,
  title={Quantum cryptography based on Bell’s theorem},
  author={Ekert, Artur K},
  journal={Physical review letters},
  volume={67},
  number={6},
  pages={661},
  year={1991},
  publisher={APS},
  doi = {https://doi.org/10.1103/PhysRevLett.67.661}
}

@article{giovannetti2011advances,
  title={Advances in quantum metrology},
  author={Giovannetti, Vittorio and Lloyd, Seth and Maccone, Lorenzo},
  journal={Nature photonics},
  volume={5},
  number={4},
  pages={222--229},
  year={2011},
  publisher={Nature Publishing Group UK London},
  doi = {https://doi.org/10.1038/nphoton.2011.35}
}

@book{amari2016information,
  title={Information geometry and its applications},
  author={Amari, Shun-ichi},
  year={2016},
  publisher={Springer},
  doi = {https://doi.org/10.1007/978-4-431-55978-8}
}

@article{lambert2023classical,
  title={From classical to quantum information geometry: a guide for physicists},
  author={Lambert, J and S{\o}rensen, ES},
  journal={New Journal of Physics},
  volume={25},
  number={8},
  pages={081201},
  year={2023},
  publisher={IOP Publishing},
  doi = {10.1088/1367-2630/aceb14}
}

@article{audenaert2007discriminating,
  title={Discriminating states: The quantum Chernoff bound},
  author={Audenaert, Koenraad MR and Calsamiglia, John and Munoz-Tapia, Ram{\'o}n and Bagan, Emilio and Masanes, Ll and Acin, Antonio and Verstraete, Frank},
  journal={Physical review letters},
  volume={98},
  number={16},
  pages={160501},
  year={2007},
  publisher={APS},
  doi = {https://doi.org/10.1103/PhysRevLett.98.160501}
}

\end{document}